\documentclass[12pt,preprint]{aastex}

\shorttitle{Radio AGN Feedback}
\shortauthors{De Young}

\begin{document}


\title{How Does Radio AGN Feedback Feed Back?}


\author{David S. De Young\altaffilmark{5}}
\affil{National Optical Astronomy Observatory, Tucson, AZ 85719}
\email{deyoung@noao.edu}



\begin{abstract}
The possible role of radio AGN "feedback" 
in conventional hierarchical cosmological models has become widely
discussed.  This paper examines some of the details of how such feedback
might work.  A basic requirement is the conversion of radio AGN outflow
energy into heating of the circumgalactic medium in a time comparable to
the relevant cooling times. First, the class of radio AGN relevant to this
process is identified as FR-I radio sources. Second, it is argued via 
comparisons with experimental data that these AGN outflows are strongly
decelerated and become fully turbulent sonic or subsonic flows due to 
their interaction with the surrounding medium.  Using this, a three-dimensional
time dependent calculation of the evolution of such turbulent MHD flows is
made to determine the time scale required for conversion of the turbulent
energy into heat.  This calculation, when coupled with observational data,
suggests that the onset of heating can occur $\sim 10^{8}$ yr after the 
fully turbulent flow is established, and this time is less than or comparable
to the local cooling times in the interstellar or circumgalactic medium for
many of
these objects.  The location of where heat deposition occurs remains uncertain,
but estimates of outflow speeds suggest that heating may occur many tens of
kpc from the center of the parent galaxy.  
Recent observations suggest that such radio AGN
outflows may become dispersed on much larger scales than previously thought,
thus possibly satisfying the requirement that heating occur over a large 
fraction of the volume occupied by the circumgalactic gas.
\end{abstract}




\section{Introduction}

The current successes of cold dark matter cosmologies, and in particular
the $\Lambda$CDM models, are well known.  These hierarchical models
begin with the microwave background fluctuations and proceed forward
to reproduce the structure in the Ly$\alpha$ forest at $z \sim 3$
(e.g., Mandelbaum et al. 2003) and continue to evolve the dark matter
structure distribution to arrive at results consistent with the 
power spectrum of the distribution of galaxies at the present epoch
(e.g., Tegmark et al. 2004), the fraction of baryons in rich
clusters (White et al. 1993), and other notable results.
These models follow the evolution of dark matter, and the accompanying
galaxies are assumed to form as gas flows into the ever growing dark
matter condensations (e.g., White and Rees 1978).
The most detailed results describing this structure evolution have
come from highly sophisticated numerical simulations that follow 
the growth of mass fluctuations in the early Universe, such as the
Millennium simulation (Springel et al. 2005). 
However, there are some well known difficulties with these models in
that they 
predict too many faint and too many bright galaxies
(e.g., Benson et al. 2003),
and in addition they predict that the largest galaxies should be blue
and star forming. 

These are serious difficulties, and they have been addressed through
the use of various assumptions about physical processes that work on
scales not resolved by the numerical simulations.  These 
"semi-analytic" models have addressed the problems at both the
dwarf galaxy and the brightest galaxy scales; the issues
addressed in this paper involve only the problems associated with
the evolution of the brightest galaxies.  The problems there are
a direct result of
the hierarchical nature of the models; 
minute fluctuations in the very early Universe
grow due to continuing dark matter and baryonic infall, 
and in the context of the models
this infall should continue to the present epoch, possibly leading to significant
star formation now in the most massive galaxies.  Thus in this picture
the most massive galaxies would be blue due to the large population of
young stars, whereas in fact the largest galaxies are generally seen as
"red and dead" at the present epoch, implying that the most active
epoch of star formation occurred earlier (e.g., Madau et al. 1996,
Cowie et al. 1996).

To address this problem, one
of the most commonly used assumptions in 
semi-analytic models is that of
"feedback" from the central regions of the galaxy, where energy in some
form is produced in sufficient quantity that it can heat the inflowing
gas, perhaps stopping or reversing the inflow, and in any case providing
enough energy to suppress star formation.  Once this has been done, the
subject galaxy will then passively evolve, 
and its aging stellar population
will become redder with time until the galaxy colors 
match the observations. Motivated by the observation of active
galactic nuclei (AGN) in many massive red galaxies, several
AGN related driving mechanisms for this feedback have been suggested.
Radiatively driven winds, AGN driven shocks and mass outflows, QSOs,
and radio sources associated with AGNs have all been proposed
(e.g., Sijacki et al. 2007).
In some instances QSOs have been referred to as "high luminosity"
AGN feedback and radio-loud AGN (or "radio AGN") have been called
"low luminosity" AGN feedback.  This may not be a meaningful distinction,
since many QSOs may well be examples of powerful radio sources that
are directed toward the observer, with their collimated emission
Doppler boosted to very high luminosities. 
The use of AGNs, and in particular radio AGNs, as sources of the
energy required to suppress star formation in feedback models has
appeal because estimates for the total energies of these objects
derived from energy equipartition arguments 
have existed for some time.  These minimum
energies (from $\sim 10^{57} - 10^{61}$ ergs)
are, in an integral sense, adequate to provide the needed
heating for AGN feedback to work.  Moreover, recent estimates of the 
total energies present in some radio sources tend to lie factors of
10 higher than the estimates based on the equipartition energy 
present in relativistic electrons and fields (B\^{i}rzan et al.
2008).

Thus the total energies available in radio AGN make these objects
viable candidates for powering feedback and for suppressing star
formation in massive galaxies, and radio AGN have been incorporated
into semi-analytic models that are based on cosmological simulations
such as the Millennium Run (e.g., Croton et al. 2006, Bower et al.
2006, Okamoto et al. 2008). 
For example, Croton et al. (2006) show that a form of radio AGN energy
input can not only reproduce the observed truncation of the bright
end of the luminosity function, but that it can also 
possibly suppress accretion
and give rise to an old, red population of elliptical galaxies.
Radio AGN feedback has also been used to explain the presence
of old "red and dead" galaxies in voids (Croton \& Farrar 2008),
where environmental effects are much less pronounced.
Finally, radio AGN feedback has also been suggested by several
authors as a means of suppressing cooling or providing reheating
in the cores of rich clusters of galaxies, thereby solving the
"cooling flow" problem and at the same time accounting for the
colors of brightest cluster galaxies (e.g, Fabian et al. 2000,
Br\"{u}ggen \& Kaiser 2002, Ruszkowski et al. 2004, 
McNamara \& Nulsen 2007).

While the integrated total
energies of radio sources are adequate, and sometimes more than
adequate (e.g., Best et al. 2006) to provide the needed feedback
in semi-analytic models, the key question of "Does it really
work?" remains.  Having enough total energy is only the first
of several conditions that need to be met.  In order for radio
AGN feedback to work, this energy contained in the outflowing
radio sources must be transferred to the surrounding medium.
In order to suppress star formation, this energy transfer has
to be fairly efficient, and equally important, 
for radio AGN feedback it has to be
turned into heat, and it has to be distributed over most of
the volume of gas in and around the halo  
of the parent galaxy.  Finally,
all this has to happen in a time less than the local cooling
time or dynamical infall time.  Until all of these conditions
can be met, the effectiveness of radio AGN feedback in solving
significant difficulties with current cosmological 
models will remain in question.  This paper provides a calculation
of one way this energy transfer process can work. 
Section II presents an overview of the
physical processes involved in radio AGN feedback, while
Section III describes a calculation of the energy transfer during
the interaction of a radio source with its environment that
covers up to 5 orders of magnitude in scale. 
The results of this calculation are given
in Section IV, and the conclusions are found in Section V.

\section{Physical Processes in Radio AGN Feedback}

A fundamental property of radio AGN that has a major impact on
the issue of feedback is the bipolar nature of the energy
outflow.  Except for low energy "compact radio sources", all
radio AGN show this morphology. The degree of collimation can
be moderate, with opening angles of the flow of order twenty degrees
or so, or it can be extreme, as in the case of the very energetic
and highly collimated flows associated with high luminosity 
radio sources and quasars.  This two-sided collimated outflow
is the principal obstacle to radio AGN feedback being an
efficient and effective mechanism for energizing and heating 
(or blowing away) the
gas in the central regions of the parent galaxy, since the
heating and suppression of star formation must occur throughout
the central region and not just along the two narrow cones in
the vicinity of the radio jets.  Thus for feedback to be effective,
some processes must be found that transfers energy from the jet to
the surrounding medium, converts that energy to heat, and 
distributes it over most of the volume in the central regions
of the parent galaxy.  Some possible mechanisms 
for this energy transfer have been
previously suggested, such as shock waves from highly supersonic
outflows (Br\"{u}ggen et al. 2007, Graham et al. 2008) 
or sound wave dissipation from less energetic
events (Sanders \& Fabian 2007, 2008).  
These will be considered in more detail later, 
but in general shock waves are one-time events that rapidly 
slow and dissipate in the direction transverse to jet motion, 
while sound waves may or may not carry enough energy to heat large volumes
of gas through their eventual dissipation.  More importantly, the
process of energy deposition by the sound waves in a magnetoionic medium
is not clearly understood in this case, especially with regard to
the dissipation lengths.

Clarification of possible
feedback processes can be found from the luminosities, morphologies and
demographics of the radio AGN outflows.  Most extended radio sources
can be placed in one of two categories first proposed by 
Fanaroff \& Riley (Fanaroff \& Riley 1974). These two classes, FR-I and
FR-II, are characterized by different luminosities and by very different
radio morphologies. 
High luminosity FR-II objects are generally characterized by
two very highly collimated jets of emission extending over large distances,
typically hundreds of kpc, with the jets terminating in regions of high
surface brightness ("hot spots") surrounded by lower surface brightness
lobes which often have a "swept back" or "bow shock" appearance
(e.g. Ferrari 1998, De Young 2002).
The morphologies of the lower luminosity FR-I objects are dramatically
different, with the highest surface brightness near the
nucleus of the parent galaxy and a gradual dimming of brightness along two
moderately collimated jets with much wider opening angles than seen in the
FR-II objects.  These outflows are also often bent or distorted and display
a meandering morphology as the flow moves outward.  This pattern of radio
emission is very suggestive of the slow, subsonic flow associated with plumes
and turbulent flows, and it was suggested many years ago (e.g., De Young 1981,
Bicknell 1984)
that these radio sources were characterized by significant interaction with
the surrounding ISM or IGM that resulted in mass entrainment and a deceleration
of the AGN outflow as it moves away from the nucleus.  It may also be that the
original outflow speeds in the FR-I objects are less than those existing in
FR-II sources, but in any case the current modeling of these objects usually 
involves deceleration 
to subsonic flow, presumably mediated by mass entrainment from the
ambient medium (e.g., Laing \& Bridle 2002, 2004).

The second key point in considering radio AGN feedback comes from the
demographics of FR-I and FR-II radio sources.  It has been known for some
time that the local space density of extended radio sources is dominated by
FR-I objects (e.g., Ledlow \& Owen 1996, Owen \& Ledlow 1997, 
Jackson \& Wall 1999, Willot et al. 2001), where the local ($z < 0.1$)
density of FR-I objects is roughly $10^{2}$ times that of FR-II radio
galaxies.  The local density of FR-I radio sources, and a calibration point
for constructing radio luminosity functions (RLF) is about 400 Gpc$^{-1}$
(Rigby, Best \& Snellen 2008).  Due to the rapid decline in the detection of
the fainter FR-I objects with distance, the dependence of the FR-I/FR-II ratio
with redshift is less clear, but recent surveys and  
modeling of the RLF out to moderate redshifts
of less than 1 suggest that the space density of FR-I objects clearly 
dominates in the local universe (Willot et al. 2001, Cara \& Lister 2008,
Donoso et al. 2009).
Thus to a very good approximation the "generic" radio AGN outflow in
the nearby universe can be taken to be an FR-I radio source. This may
also be the case at moderate redshifts as well, since the previous picture
of a constant FR-I space density with cosmic epoch coupled with a rapidly
evolving FR-II space density may be overly simple (e.g., Rigby, Best \&
Snellen 2008).  In any case it is clear that "radio AGN feedback" for
redshifts less than one can be taken as "FR-I AGN feedback", and henceforth
the focus here will be on FR-I objects as sources of radio AGN feedback.  

A third observational result that constrains the feedback process comes from
radio and x-ray observations of radio sources in rich clusters of galaxies.
For the first time since their discovery over thirty years ago, it is now
possible to carry out calorimetry of some radio source outflows by using
estimates of the
$pV$ work done by the outflows in inflating cavities observed in the ICM.
(e.g., B\^{i}rzan et al. 2008, McNamara \& Nulsen 2007 and references therein.)
The role of such cavities as being a signature of radio AGN feedback in clusters
has been discussed elsewhere, and the relation of this phenomenon to more
general FR-I outflows will be discussed at the end of Section 4. 
The focus here is on the general problem of radio AGN feedback,
both in and out of clusters and groups of galaxies, and the relevant result from
the radio source calorimetry in clusters is that it reveals outflow energies
that are generally in excess of the equipartition values that were commonly
assumed in the past (B\^{i}rzan et al. 2008).  These $pV$ values are often an
order of magnitude greater than the total energies contained in relativistic
electrons and magnetic fields in minimum energy equipartition, and they are
consistent with most of the energy residing in the kinetic
energy flux of the outflow (e.g., De Young 2006). 

With these three observational constraints - the meandering plume-like
morphology of FR-I radio sources, the dominance of FR-I objects in the
nearby radio AGN population, and the existence of much more energy in
the outflows than is required to account for the synchrotron radio
emission, the most likely paths for radio AGN feedback become more
clearly defined.  First, if these outflows are truly subsonic over
most of their length, and the morphological data are very suggestive
of this, then heating by shock 
waves is not likely to be a significant energy transfer mechanism
in these objects, nor is momentum transfer likely to form a 
nearly spherical outflowing wind that removes the ambient gas. 
Second, because there is strong evidence that
AGN outflows are relativistic on very small ($\sim$ parsec)
scales, the FR-I morphology strongly suggests significant 
deceleration of the flow.  The uniform appearance of the outflows from
the nucleus outward without any clear discontinuities 
together with the spreading of the outflow from the nucleus outward
implies that this deceleration process begins near the nucleus and
continues more or less uniformly throughout the outflow region within
the parent galaxy.  Deceleration of the flow means a loss of
momentum from the jet via a transfer of momentum from the jet to another
medium, with the most obvious candidate being the ambient gas in the
ISM of the host galaxy.  The most natural and most obvious candidate
for this deceleration process is thus entrainment of, and transfer
of momentum to, the ambient medium at the interface between the
outflowing AGN material and the ISM and galactic halo gas.

In addition to continuous deceleration from entrainment,
deceleration of the outflow from its original speed near the
galactic nucleus could entail the production of shock waves and subsequent
heating, since VLBI observations of AGN suggest very rapid outflow speeds on
the parsec and subparsec scales in the innermost regions.  However, the 
propagation of a decelerating supersonic jet into the circumnuclear gas will
produce a driven bow-shock configuration that, for uniform flow, is a one-time
event.  In addition, except for the small angular region near the head of
the jet, such shocks are generally weak, oblique, and decelerating. 
Nonetheless, they
probably occur around the initial outflow regions of the jets, and as
such they could produce localized heating in the innermost regions of the
parent galaxy, especially if the AGN outflow involves multiple outbursts.
This process may be at work in outflows in rich clusters such as 3C 84, where
deceleration is most dramatic due to the very high ambient densities.
Because the total energy involved in these shocks is much less than the kinetic
energy of the outflow, and because the focus here is not on AGN in rich clusters,
the emphasis here will be on the transport of energy by the outflow itself,
with most of the flow deceleration being due to continuous entrainment of
the ambient medium.

This entrainment process is mediated through
the non-linear development of surface instabilities, most 
notably the Kelvin-Helmholtz (KH) instability, and it is driven by
the large ($\sim 10^{43} - 10^{45}$ erg s$^{-1}$) kinetic energy 
flux of these outflows derived from the calorimetry described above.  
Analytic 
approximations, laboratory experiments, and numerical simulations
have all shown that this instability in shearing flows is
essentially inevitable.  Incompressible flows, compressible
flows, supersonic flows, MHD flows and relativistic shearing
flows all show the onset and non-linear growth of the KH
instability (e.g., Chandrasekhar 1961, Brown \& Roshko 1974,
Clemens \& Mungal 1995, Aloy et al, 1999, Ryu, Jones \& Frank 2000,
Perucho et al. 2004, 2007).
In the fully non-linear stage of development the instability 
evolves into a mixing layer dominated by large scale vortex
structures that entrain material from both sides of the layer
into the layer itself (e.g., Dimotakis \& Brown 1976).
The thickness of this layer grows as one moves down the jet from
the onset of the non-linear phase of the instability, and the
growth of the layer is nearly linear with distance along the
jet until the jet is fully infiltrated by the layer.  At that
point a different evolutionary phase begins, as is described in
Section 3.  For a plane mixing layer, which is a good 
approximation to the cylindrical mixing layer in a round jet
in its initial stages (e.g., Dimotakis et al. 1983, 
Freund et al. 2000), the initial growth is characterized by
a nearly constant opening angle that depends upon the
relative flow speeds and the fluid densities as
\begin{equation}
Tan \theta \approx (C/2)(\rho_{j}/\rho_{a})^{-\eta}M^{-1}, 
\end{equation}
where $\rho_{j}$ and $\rho_{a}$ are the densities in the outflow
and the ambient medium respectively, and $M$ is the Mach number of the
outflow relative to the sound speed in the ambient medium
(e.g., Brown \& Roshko 1974, Dimotakis \& Brown 1976, De Young 1993).
Empirical data suggest $C \approx  0.16$ and $\eta \approx  0.5$.
Within this growing mixing layer the entrainment proceeds 
through the initial production of large scale eddies whose diameter is
roughly equal to the thickness of the layer, followed by the
development of finer scale structures within the eddies as well
as merging of the larger scale structures.  In other words,
the mixing layer becomes fully turbulent.  Moreover, the thickness
of this turbulent mixing layer grows as one moves downstream along
the jet, and eventually the turbulence will penetrate throughout
the entire volume of the outflow.  From Eq. 1, the distance along
the jet from the onset of the non-linear mixing layer to the point of 
complete infiltration of turbulence into the interior of the jet
can be approximated by
$$L_{SAT} \sim (2/C) R_{o} M (\rho_{j}/\rho_{a})^{\eta},$$
where $R_{o}$ is the radius of the jet at the onset of the mixing layer.
Beyond this point the flow within the jet is fully turbulent; many
experimental examples of such turbulent, decelerated jets can be found
(e.g., Dimotakis et al. 1983, Mungal et al. 1992), 
and within the limits of resolution the outflow from
FR-I objects such as 3C 31 is indistinguishable from these experimental
results.
Thus the FR-I demographics, when combined with 
the evidence for uniform deceleration in their
AGN outflows, the universality of shear driven surface instabilities
that entrain mass and provide momentum transfer through the growth
of turbulent mixing layers, and the morphological similarities
between FR-I outflows and subsonic turbulent plumes all provide
evidence that the most common
energy outflow associated with radio AGN feedback is that of 
diverging, subsonic, fully turbulent plumes.

This argument is given further support from a comparison of high
resolution radio observations of FR-I outflows with 
experimental results concerning fully turbulent round jets propagating 
into a uniform medium.  There is
considerable evidence that such jets are self-similar
flows (e.g., Pope 2000, Hussein et al. 1994), and
that these flows all have the same opening angle once self-similarity
is established.  This angle is about 23-24 degrees, and an examination
of radio observations of the FR-I objects 3C 31 and 3C 296 (Hardcastle
et al. 2005, Laing et al. 2008) shows that both the north and south
jets in these objects are very well enclosed by an opening angle 
of about 23 degrees ($tan(\theta/2)=0.20$).  
Once the outflows show sudden divergence or bending this
is no longer the case, but in the inner regions of the flow where 
the ambient conditions are likely to be fairly constant or slowly 
varying, the agreement with self-similar turbulent jets is remarkable.
Elements of this picture
have been suggested before, 
and the first numerical simulation of mass
entrainment in astrophysical jets was performed many years
ago (De Young 1986); there has since been considerable 
development of phenomenological models of these flows
(e.g., Laing et al. 1999, Laing \& Bridle 2002, 2004, Wang et al. 2009).

Because magnetic fields are present in both the AGN outflows and in the
ambient medium, possible inhibition of the mixing layer growth by magnetic
effects might occur, especially for "favorable" field geometries where 
${ \bf B}$ is parallel to ${ \bf v}$.  However, it is clear that in the
context of MHD outflow models, the average kinetic energy density must 
be much greater than the average magnetic energy density in order for the
AGN outflow to occur. (The opposite case of nearly mass-free "Poynting"
jets is discussed in the last section.)  In the MHD case 
the concern is then that very
local B field amplification due to turbulence could lead to local field
energies sufficient to damp the development of the mixing layer.  
In this case, detailed 3D
numerical MHD simulations of the development of the K-H instability and
its fully non-linear growth (e.g., Ryu et al. 2000) show that the flow
remains "essentially hydrodynamic".  Small scale field amplifications do
occur, and at late times the amplitudes are comparable to the flow energy
densities, but these areas are very localized and do not inhibit the 
growth of turbulent flow in the mixing layer on the larger scales of 
energy injection into the turbulence.  It is important to note the 
difference between these dynamic, driven outflows and the much slower,
passive evolution of buoyant "bubbles' in an intracluster medium, where
large scale ICM fields may have an influence on the bubble evolution.

This characterization of radio AGN feedback as turbulent FR-I outflows
provides the initial conditions for finding
processes that can transform this outflow into 
widespread heating of the ambient medium. 
A major advantage of turbulent flows is that the 
ultimate fate of fully developed turbulent flows is conversion
of the turbulent energy into heat.  Thus if mass entrainment and
the generation of turbulence is the most efficient coupling of
the AGN outflow to the ambient medium, then the the natural 
evolution of that turbulence will provide the most effective
heating of the gas in the galaxy by radio AGN outflows.
A central issue is how long this conversion
takes.  The next section provides a quantitative estimate of this
time through a calculation of 
the evolution of a driven 
turbulent flow from its onset until its dissipation.

\section{Evolution of Turbulent AGN Outflows}

\subsection{Formalism}
The turbulent flow that describes the radio AGN outflow can be
recast in a form common to many turbulent flows.
On some large scale comparable to a defining scale of the problem
(e.g., the radius of the outflow) energy is injected in the form
of turbulent eddies.  This process, as verified by experiment,
involves a large scale ingestion of material on both sides of
flow boundary in a "gulping" mode by the largest eddies (e.g.,
Dimotakis \& Brown 1976),
followed by a rapid generation of finer and finer structure
and mixing 
within this large scale as the flow evolves to smaller scales.
In terms of the evolution of the energy spectrum of the turbulence,
this can be described as energy injection over a small range of
wavenumber $k$ followed by a loss free cascade of energy to larger
wavenumbers.  For homogeneous and isotropic hydrodynamic 
turbulence this process results in the standard power-law 
Kolmogorov spectrum with $E_{v} \propto k^{5/3}$.  (For MHD
turbulence the power-law exponent is less well defined but is
generally somewhat smaller.) 
The dissipation
of the turbulent kinetic energy into heat occurs at and above
some wavenumber $k_{d}$ which generally lies many orders of 
magnitude above the energy injection scale $k_{I}$.  The 
Kolmogorov formalism describes an equilibrium state where the
energy input at $k_{I}$ is exactly balanced by the dissipation
at $k_{d}$.  The problem being addressed here is not as simple,
since what is needed is a calculation of the time required for
the flow to reach the dissipation range at $k_{d}$ after the
injection has begun at $k_{I}$.  Because the evolution of the
turbulent flow involves the non-linear interaction and energy
transfer among turbulent scales of many different sizes, both
upward and downward in $k$ space, the evolution must be 
calculated explicitly.  In addition to the time dependent nature
of the flow, a major hurdle is the many orders of magnitude in
scale between the size of the large eddy injection range
(tens of parsecs or more) to the energy dissipation scale due
to effective viscosities in ionized plasma (possibly meters
or less).

Some simplifying assumptions can be made. 
If the overall outflow of the FR-I AGNs is 
subsonic or transonic relative to the ambient medium, then
the flow can be regarded as generally incompressible.  In
addition, in a frame comoving with the mean local outflow speed 
the turbulence can be treated as locally homogeneous and weakly
isotropic (in that helicity is allowed)
to a good approximation.  Under these conditions a
technique may be used that calculates the three dimensional
time dependent evolution of turbulence, both hydrodynamic and
magnetohydrodynamic, over several orders of magnitude in
spatial scale.
The original method was devised to treat purely hydrodynamical
turbulence (e.g., Orszag 1970, 1972) but was subsequently generalized
to include magnetic fields in the MHD approximation (Pouquet
et al. 1976, Pouquet et al. 1978, De Young 1980).
This approach takes moments of the Fourier transforms of the 
equations of mass, momentum and energy conservation in order
to obtain equations that describe the time evolution of second
moments of velocity and magnetic field, i.e., the kinetic energy
and magnetic energy, as a function of time and spatial scale.
The system of moments is closed by using a quasi-normal approximation
that assumes 
the fourth order moments are related to the second order
moments in the same way as is the case for a Gaussian distribution of
Fourier modes.  This "quasi-normal" approximation permits closure of
the system of equations at second order and results in
differential equations that describe the temporal and spatial
evolution in Fourier space of the square of the fluid velocity
(i.e., kinetic energy per unit mass) and the magnetic energy per
unit mass.  The calculation uses a complex form of eddy viscosity that
employs a Markov cutoff in "memory" of the eddy interactions and 
includes the non-linear transfer of energy
among turbulent eddies of different sizes, including triads of eddies
with differing wavenumbers.  Some additional details are found in the
Appendix.  
The principal advantages of this eddy damped,
quasi-normal Markov (EDQNM) method are its ability to calculate the
time evolution of the turbulence over very large spatial ranges and
its relatively modest computational requirements, especially when
compared to the more straightforward methods of direct numerical
simulation.  The technique is is wide use, and its accuracy
has been verified by comparison with both direct numerical simulations
and with laboratory experiments (e.g., Gomez et al. 2007, Vedula et al.
2005, Turner \& Pratt 2002, Lesieur \& Ossia 2000, Staquet \&
Godeferd 1998).

Details of this technique 
can be found in Orszag (1977), Lesieur (2008), and Pouquet et
al. (1976) and De Young (1980) for the MHD generalization.
As discussed in the Appendix,
evolution of the turbulence is followed by solving a series of 
integro-differential equations of the general form
\begin{equation}
(\partial/\partial{t} + 2\nu k^{2})E(k) = F(k) + 
\int T[k,p,q,E(k,p,q)\nu]dpdq.
\end{equation}
Here $E$ is the Fourier transformed magnetic or kinetic energy at
wavenumber $k$, $\nu$ is a dissipation constant (either viscosity
or magnetic diffusivity), $F(k)$ is the forcing function or driving
term, which in this case arises from the surface instabilities at
the jet-ambient medium interface that create
the large scale turbulent eddies driving the turbulent flow.
The term $T$ represents the non-linear transfer terms in the flow
which mediate the transfer of magnetic and kinetic energies over all
wavenumbers; i.e., $T$ describes the processes that create the
cascade of energy from large scale to small scales, the amplification
of magnetic fields by fluid turbulence, and the inverse cascades, if
any, of energy from small scales to larger scales. (Which in general
is significant only in the presence of flows with net helicity.)
In general all of $E$, $F$, and $T$ are time dependent.  The 
explicit form for $T$ can be complex because it describes the 
interaction of hydrodynamic turbulent eddies with each other, the
interaction of hydrodynamic turbulence with any magnetic fields
and the resulting possible magnetic field amplification, 
the effects of eddy viscosity, and the
self interaction of magnetic turbulent structures with each other.
The complete MHD forms of $T$, including the effects of 
helicity ({\bf $v\cdot\nabla\times v$} and its magnetic counterpart
{\bf $b\cdot\nabla\times b$}) are given in 
De Young (1980).

In the present case magnetic effects are not of major concern.
However, a very small, dynamically unimportant magnetic field
is included for completeness; its initial energy density is
$10^{-2}$ that of the turbulent kinetic energy density.  It
could in principle play a role in the dissipation process at
small scales, but to include these effects would require a
knowledge of the detailed field reconnection and dissipation
mechanisms, which in turn would require a specification of 
field geometries and strengths as a function of scale length.
Since these are unknown, the magnetic diffusivity is chosen to
be consistent with the hydrodynamic dissipation as described 
below.  Magnetic field amplification to equipartition with the
hydrodynamic energies will occur on very small scales, but in
the absence of any helicity this will have no effect on the
evolution of the turbulence on the scales of interest here.

The problem is specified by the injection of kinetic
turbulent energy at some rate $F(k)$ at a scale defined as the
energy range, and the evolution of the turbulent energy spectrum
is then followed as the energy cascades to smaller and smaller
scales.  A natural unit for length is that corresponding to the
energy injection scale, $L_{o} \propto 1/k_{I}$, and a corresponding
time scale comes from the largest eddy turnover time; i.e.,
$t_{o} \approx L_{o}/v_{o}$, where $v_{o}$ is the mean velocity
of the turbulent flow.  Because the turbulence is driven by shear
instabilities in the mixing layer between the AGN outflow and the
surrounding medium, $v_{o}$ is comparable to the local jet outflow
speed.  The calculations here are basically scale free.  This is
because the actual dissipation scale is very many orders of
magnitude smaller than the energy injection scale, and inclusion
of the energy scale, which is necessary, means that even with
this technique the calculation cannot span the scale lengths from
energy injection to dissipation. The dissipation scale is given by
$k_{d} \approx k_{I}R^{3/4}$, where $R$ is the Reynolds number
appropriate to the large scale flow and $k_{I}$ is the wavenumber
corresponding to the energy injection scale.  The hydrodynamic
Reynolds number is $R = <v_{I}^{2}>^{1/2}/\nu k_{I}$, where 
$<v_{I}^{2}>^{1/2}$ is the average velocity at scales $k_{I}$;
i.e., $v_{I} \sim v_{o}$.
The magnetic Reynolds number is just this expression with the
magnetic diffusivity $\lambda$ substituted for the kinetic 
viscosity $\nu$. For a hot ($\sim 10^{7}$K) rarefied ($n \sim
10^{-4}$) plasma $\nu \sim 10^{-4}$ (e.g., Spitzer 1962),
and energy injection scales of tens of parsecs and sonic
or subsonic flow speeds of $10^{2} - 10^{3}$ km/sec  
imply ten or more orders of magnitude in scale between the
energy injection scale and the dissipation scale.  Even though
the equations of the form (2) above are solved numerically
on a logarithmic scale in $k$ space, a solution that spans the
complete scale from $k_{I}$ to $k_{d}$ is not feasible, especially
in view of a causality condition discussed below.  Thus the actual
values of the dissipation constants $\nu$ and $\lambda$ never
enter into the calculation, and no inherent scales are present.
This means that application of the results can be applied to
many different spatial and time scales, and this characteristic
will be used in estimating the time for the turbulent energy
flow to actually reach the regime where energy is dissipated into
heating the ambient medium.

Because the wavenumber range included in the calculations is less
than the complete range to $k_{d}$, some mechanism has to be used
at the upper end $k_{top}$ of the calculated wavenumber range to simulate
the continuation of the loss free cascade of energy to wavenumbers
beyond $k_{top}$.  If this is not done, unphysical reflection of 
turbulent energy from the upper end of the calculated wavenumber
range can occur.  A standard method for treating this is to use
artificially large values of the dissipation constants $\nu$ and
$\lambda$.  These are adjusted so that the resulting 
effective values of
$k_{d}$ are somewhat less than $k_{top}$ so that no spurious 
energy reflections are seen.  
The differential equations of the form given by Eq. 2 are solved
numerically on a discrete grid in $k$ space, and in order to avoid
non-physical energy fluctuations, a time step control similar to a
Courant condition must be used.  The value of the time step in the
integration is chosen to be less than the minimum of 
$[1/(v_{I}k_{top}), 1/(k_{top}^{2} \nu)$.  This ensures the correct
behavior at high wavenumbers; it also illustrates why extending the
calculated wavenumber range beyond about 5 orders of magnitude is not
practical.  Because the calculation is scale free, the obvious scaled
values to choose are $k_{I}=1$ and $v_{I}=1$; once the physical dimension
and speed corresponding to these are chosen, the time step is determined 
and the problem is well defined.

The scale-free nature of the calculation allows a simple estimate of the
time required to actually reach the true dissipation range once a
calculation has reached equilibrium over the more limited wavenumber
range described above.  Suppose a calculation is made of the evolution
of the turbulence spectrum between some $k_{1}$ and $k_{2}$ using a
$\delta$-function energy input at $k_{1}$.  Let $k_{2}$ be some large 
multiple $M_{1}$ of $k_{1}$, about 100, and suppose the spectrum reaches
a steady state at some time $t_{1}$, where $t$ is measured in units
of $t_{1,o} = \ell_{1}/v_{1}$ with $\ell_{1} \sim 1/k_{1}$;
$t_{1,o}$ is 
roughly the large scale eddy turnover time at $k_{1}$, and $v_{1}$ 
is the mean flow speed of the turbulence on this scale.
Hence $t_{1}$
can be measured in units of the turnover time and thus $t_{1} \approx
N_{1} t_{1,o}$.  Because  
it is scale free, the same calculation also applies to driving turbulence
between $k_{2}$ and $k_{3} = M_{2} \times k_{2}$ with 
a $\delta$-function at $k_{2}$.  In this case a steady state is 
reached in some time $t_{2}$ where now $t$ is measured in units 
of $t_{2,o} = \ell_{2}/v_{2}$ with $\ell_{2} \sim 1/k_{2}$.  Similarly,
the calculation can be made between $k_{3}$ and $k_{4} = M_{3} \times k_{3}$,
and so on until some $k_{n}$ is reached where $k_{n} \approx k_{d}$.
Then the total time required for the turbulence to cascade from $k_{1}$
to the dissipation range is just $t_{d} = t_{1}+t_{2}+t_{3}+...+t_{n}$.
But $t_{1} = N_{1}t_{1,o} = N_{1}\ell_{1}/v_{1}$ and similarly for   
the other $t_{i}$.  But in this scale-free calculation, 
$M_{1}=M_{2}=...=M_{n}=M$, and thus $N_{1}=N_{2}=N_{3}=...N_{n}=N$, and
so $t_{d} = N(\ell_{1}/v_{1} + \ell_{2}/v_{2}+...+\ell_{n}/v_{n})$.
To a first approximation all of the $v_{i}$ are equal, though it 
could be argued that as one goes to smaller and smaller scales the
relevant values of $v_{i}$ decrease due to an increasing effective
viscosity, especially in the presence of magnetic fields.  By setting
all $v_{i} = v_{1}$ a lower limit on the value of $t_{d}$ is 
obtained, and this is the most constraining value for the issue of
concern here.  Then $t_{d} \approx (N/v)\Sigma \ell_{i}$, and since
$\ell_{i} \sim 1/k_{i}$ the time for the turbulent flow to reach
the dissipation range and begin to heat the ambient medium is
\begin{equation}
t_{d} \simeq (N\ell_{1}/v)[1 + 1/M + 1/M^{2} + ...+ 1/M^{n-1}].
\end{equation}
Thus a good approximation of the time for the turbulent cascade
to reach the dissipation stage is just the number $N$ of large scale
eddy turnover times that are required for the establishment of a
steady state turbulent cascade over a factor 
of about $M \approx 100$ in scale 
and beginning with the scale of the large scale energy injection.
i.e., $t_{d} \approx N\ell_{1}/v_{1}$.

\subsection{Initial Conditions}

In this model the energy injection occurs on the scale of the
largest eddies when the outflowing jet first becomes fully
turbulent from outer boundary to centerline.  In the limiting
case the energy injection spectrum could be taken as a delta
function centered on $k = k_{I} =1$, and the energy injection
function in Eq. 2 would have the 
form $F(k) = F_{v}\delta(k=k_{I}$).
Since the goal of the calculation is to determine the time 
required for the turbulence originating on a scale $k_{I}$ to
cascade down to the dissipation range, the injection spectrum
should be some sharply peaked function centered on $k_{I}$.
A narrow Gaussian form may be more cosmetically appealing but
will be seen to have little effect on the outcome of the
calculation.  The value of the constant $F_{v}$ controls the
amplitude of the energy injection spectrum, and because the
calculation is looking at the minimum time required for the
turbulent cascade to reach the dissipation range, values of
$F_{v}$ near the maximum allowed are of interest.  The units of
$F_{v}$ are (kinetic) energy per unit mass per unit time, and
in the dimensionless units used here, $F_{v}$ should be some
fraction of $v_{I}^{3}/R_{t}$, where $R_{t}$ is the value of the
jet radius at which the flow becomes fully turbulent throughout
its volume.  The maximum value of $F_{v}$ is then equal to one. 
A "reasonable" choice might be to set an average
large scale eddy size to half $R_{t}$ and the similarly for some
average rotation speed of those eddies as $v_{I}/2$, which
gives a value of $F_{v} = 0.25$ in dimensionless units where
$v_{I} = 1$ and $R_{t} = 1$.
In fact, it will be seen that
values of $F_{v}$ between 1.0 and 0.1 yield roughly the same
results.  Introducing a time variation in $F_{v}$ is easily
done, but such variations could only be less than the maximal
value and thus defeat the purpose of finding minimum equilibration
times.  Hence $F_{v}$ will be kept constant.

A major virtue of the formalism used here is that the calculation
is carried out with logarithmic intervals in $k$ space, and though
this introduces some complications to ensure that energy transfer
among structures in neighboring regions of wavenumber space is not
incomplete, it allows the calculation to cover many orders of
magnitude in scale and thus permits the development of a turbulent
cascade through the loss-free inertial range.  In general the
calculations here cover almost four orders of magnitude in scale in  
three dimensions, which would require a major computational effort
in a direct numerical simulation.  Most calculations used a 
wavenumber range of $k \in [0.015,128]$, while more extensive
runs extended the upper limit to $k = 1024$.
The kinetic energy in the EDQNM formalism must be positive definite,
so the initial energy spectra were set equal to a constant but
negligible value of $10^{-4}$ in scaled units at all wavenumbers. 
This is three to four orders of magnitude below the injection energy
and has no effect on the evolution of the energy spectrum resulting
from the turbulent mixing layer.  The energy levels at the upper and
lower ends of the wavenumber range were not fixed after initialization.
In practice there is no significant inverse cascade of energy in the
absence of helicity, so the value of $E_{v}$ at the low end of the
wavenumber range (largest scale) does not evolve from its initial
value.  The energy at the largest wavenumber $k_{top}$ begins to fall 
immediately due to the very rapid energy transfer on the smallest scales, 
but this value always remains non-zero. 

\subsubsection{Parameter Space}

The major free parameters in these calculations are the shape and
amplitude of the energy injection spectrum $F_{v}(k)$.  Other parameters
that can be varied are the wavenumber range, the length of time covered
by the calculation, the value of the viscosity to test for any influence
on the spectral evolution, and the density of points in wavenumber space.
For most of the calculations the shape of $F_{v}(k)$ was taken to be
a delta function centered at the energy injection scale $k_{I}$, as
discussed above.  In some runs this shape was changed to a Gaussian
distribution in the log of  
injection energy sharply peaked at $k = k_{I}$, and with
$\sigma = 0.2$ in log$k$.
The values of $F_{v}$ ranged from the maximum of 1.0 to
0.01, with the most common value being 0.25 as described previously.
The normalization constant for the Gaussian runs was chosen so that
the peak of the distribution had the same amplitude at $k=k_{I}$ as
did the delta function energy injection, thus providing a comparable
energy injection rate. The scaled wavenumber range was in most cases
from 0.015 to 128, with a few runs being made over the range
$k \in [0.015, 1024]$. Values of the time evolution of the turbulent
energy spectra were calculated at equally spaced logarithmic 
intervals in the $k$ domain; some runs were repeated with the
density of points increased by a factor of two to see what effect,
if any, would result from changing the size of $\Delta k$.
As described above, values of the viscosity $\nu$ were chosen so
that an effective $k_{d}$ lay near the value of $k_{top}$.
Various small changes in $\nu$ were made in some runs to find the
smallest value of $\nu$ that would not produce any "feedback" from
the upper end of the $k$ range into the larger scales.  Such
unphysical inverse cascades were immediately apparent by the growth
or "bunching up" of elevated $E_{v}$ values at the top end of the
wavenumber range.  Adequately high dissipation to ensure only a
direct and unimpeded cascade to smaller scales (large $k$) was always 
shown by a rolloff in values of $E_{v}$ from a power law at the
highest wavenumbers.  The timestep was determined for each run via 
the effective Courant condition described earlier; increases in 
$\Delta t$ much above this value immediately result in zero or 
negative values of $E_{v}$ at the highest wavenumbers.

\section{Results}

The objective of these calculations is a determination of the time
elapsed from the onset of driven turbulence on some large scale to
the transformation of the turbulent energy into heat at the dissipation
scale.  This first occurs when the turbulent cascade reaches an 
equilibrium or steady state condition with energy flowing in at $k_{I}$
and being dissipated at $k_{d}$ after transit through a loss free
cascade to ever smaller scales.  Establishment of this equilibrium 
is an asymptotic process, hence the time required for equilibrium
will be set at that time when any changes in the energy spectrum per
unit time are less than a few percent.  This will give a
minimum time for establishment of equilibrium and will result in the
minimum time for feedback into heat to occur. 

The approach to equilibrium is slightly complicated because there is
no well defined "front" of energy marching down to smaller scales into
a constant background of very low turbulent energy.  Instead, as soon
as the calculation is begun, energy begins to flow to higher 
wavenumbers, as it should.  This means that the initial constant and
very small value of $E_{v}$ set at the beginning shows a decay at
large $k$ before the energy flow from the injection region arrives at
the smallest scales.  Hence the shape of the spectrum evolves over the
\begin{figure}
\plotone{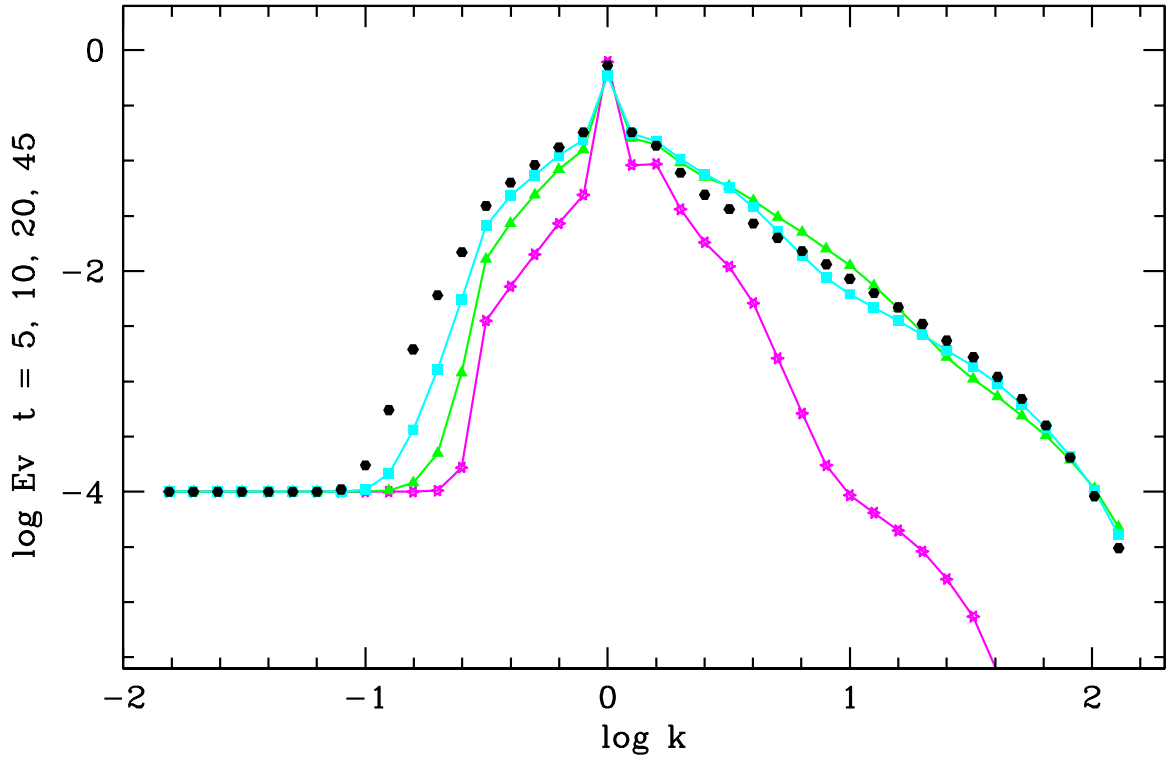}
\caption{Early time evolution of the turbulent kinetic energy spectrum.
Energy is injected with a delta function at $k=1$, and time is measured
in units of eddy turnover time at $k=1$. Magenta stars: t=5; green 
triangles: t=10; blue squares: t=20; black circles: t=45.  
The spectrum has not reached
equilibrium by $t=45$. }
\end{figure}
full wavenumber range for all times greater than zero.  Figure 1, which 
shows the early evolution of the turbulent kinetic energy spectrum as
a function of time, illustrates this behavior.  In this case the energy
injection spectrum has a 
delta function form for $F(k)$ at $k=1$, and time is
measured in units of the $k=1$ eddy turnover time, $t_{o}=R_{t}/v_{I}$. 
At early times ($t \leq 5$) the effects of the energy drain at large
wavenumber is very evident, since the spectrum in this range is reduced
far below the initial constant value of $10^{-4}$ described in 
Section 3.2.  Nonetheless, at $t=5$ the flow
of energy from the injection region at $k=1$ to smaller scales is clearly
seen, with a "front" of the energy cascade having arrived as far as 
$k \approx 10$.  By $t=10$ the effects of energy injection have reached
to scales of order $10^{-2}$ smaller than the injection region, but,
as Figure 1 shows, the spectrum has not yet reached an equilibrium state
by $t=45$.  Figure 2 shows the late time evolution of the kinetic energy
spectrum, and there it is clear that for scales smaller than the injection
scale, there is very little evolution in the spectrum at times later
than $t=60$, with little or no evolution past $t=100$.  At larger scales
there continues to be a slow growth of turbulent energy on scales of up
to 10 times the size of the injection region, consistent with previous
calculations.  However, this region is of no concern for the present
problem, and in addition the applicability of homogeneity on these scales
is less certain for the jet structures considered here.  The region $k < 1$
has no effect on small scale evolution, and in the absence of helicity
almost all the injected energy cascades to the small scale regions (e.g.,
De Young 1980).

\begin{figure}
\plotone{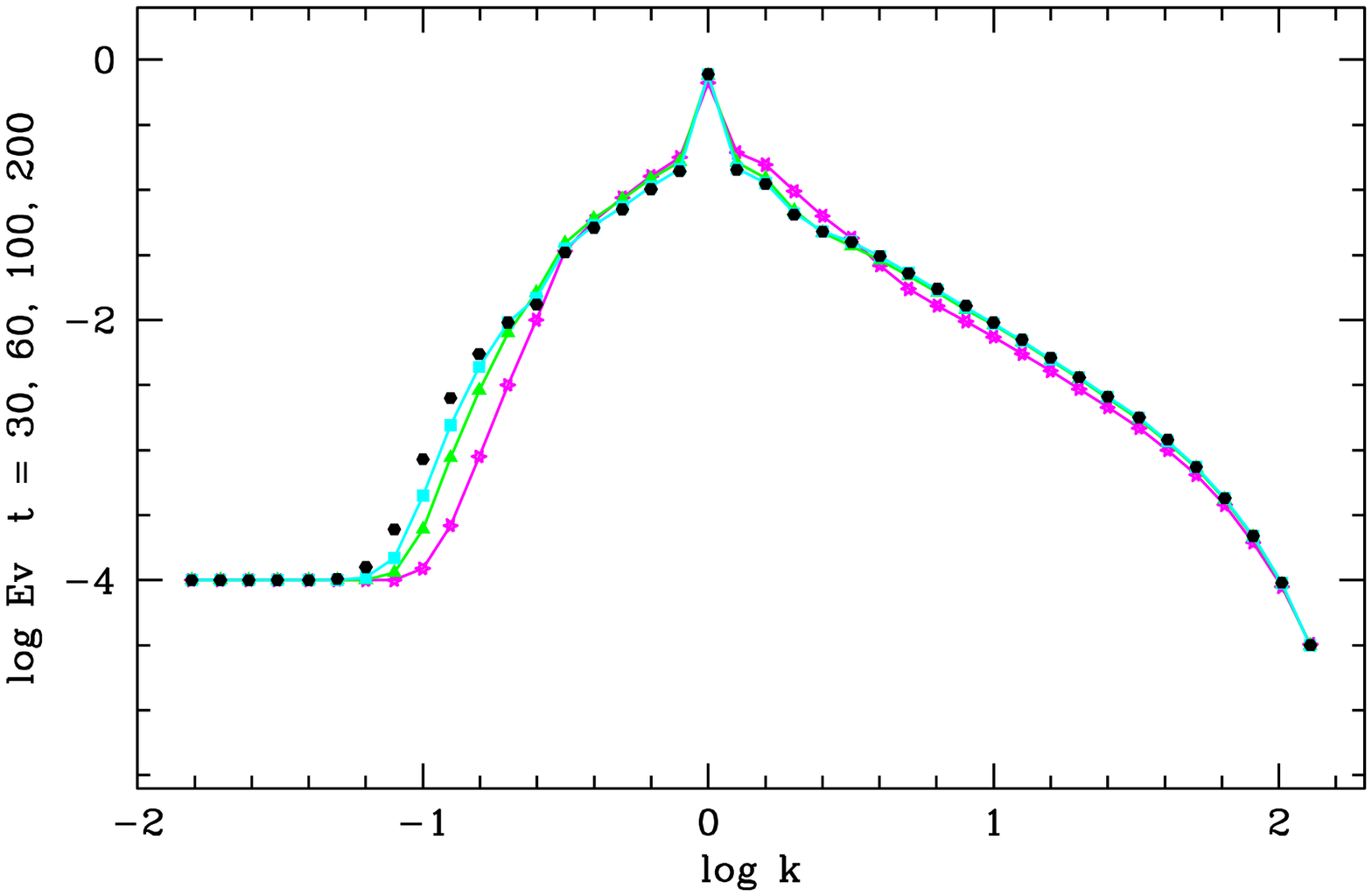}
\caption{Late time evolution of the kinetic energy spectrum shown in
Figure 1.  Magenta stars: t=10; green triangles: t=60; blue
squares: t=100; black circles: t=200.  
The spectrum has reached equilibrium by $t=200$ for all
values of $k \geq 1$.}
\end{figure}

Figure 3 shows the resultant spectrum at $t=60$ when the wavenumber space
is extended to $k=1024$, thus spanning over 5 orders of magnitude in scale.
As discussed above, this is a practical maximum for a single calculation
due to the time step constraints at the smallest scales.  Figure 3 also
shows a comparison between a $\delta$ function energy injection at $k=1$
and a Gaussian energy injection on the same scale with the normalization
discussed in the previous section.  It is clear that there is no
significant difference in the late time spectra that result from the two
different energy injection functions.  Also evident in Figure 3 is the
establishment of a clear power-law energy spectrum for times greater than
\begin{figure}
\plotone{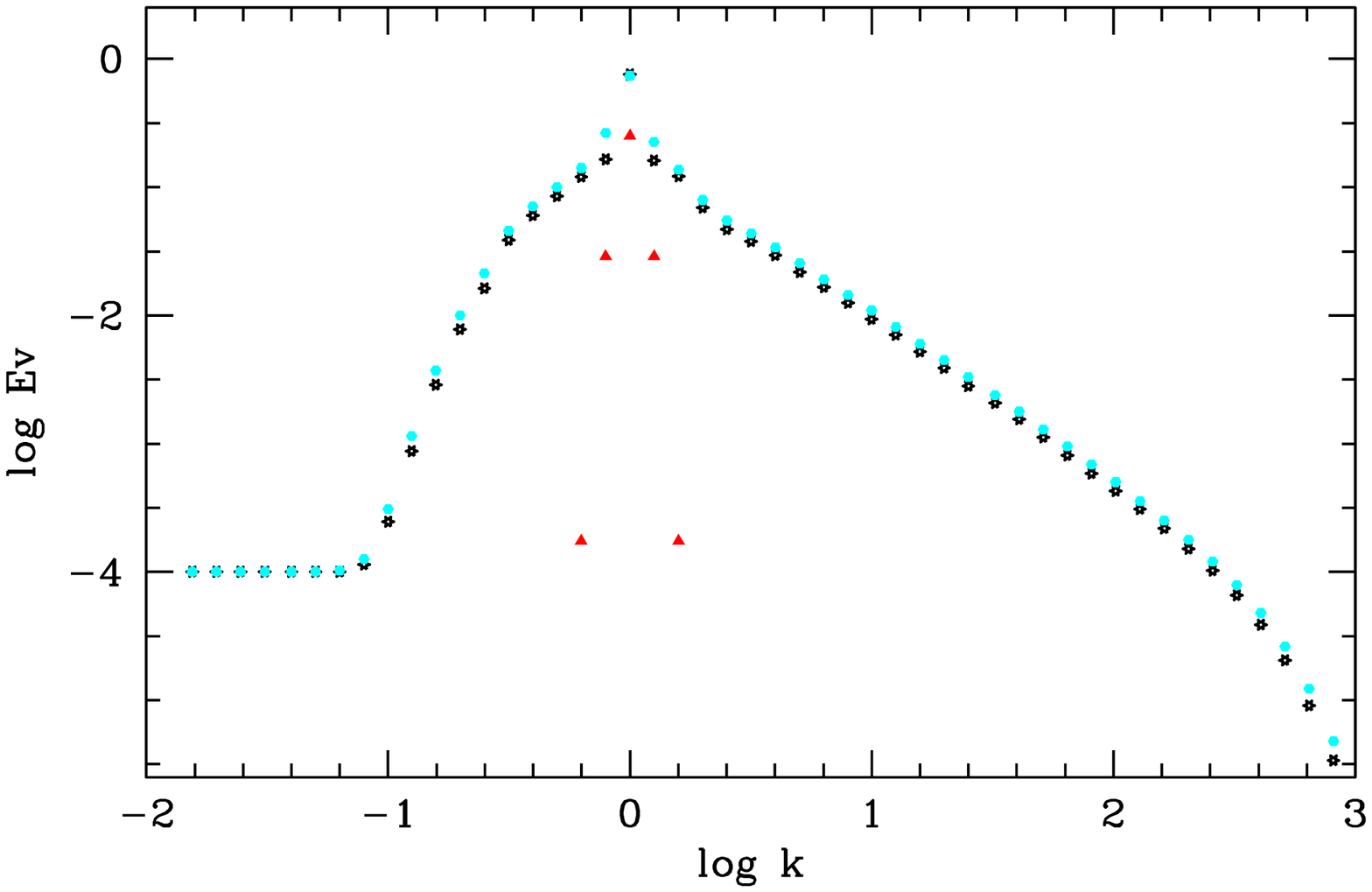}
\caption{Turbulent kinetic energy spectra at a scaled time of $t=60$ for
a wavenumber space extending to $k=1024$. Shown are the spectra
for delta function (black) and Gaussian (blue) energy injection profiles, 
with the form of the Gaussian injection spectrum also shown 
by red triangles.  No appreciable differences
exist in the overall energy spectrum for the two different injection 
profiles.}
\end{figure}
$t=60$.  Finally, Figure 4 shows the time evolution of the total (kinetic
plus magnetic) energy for the $\delta$ function injection case that covers
the range $k \in (0.015, 1024)$.  From this figure it is clear that the
turbulence has not reached an equilibrium state 
for times less than $t \simeq 70$.
However, it does seem that equilibrium is established by $t=100$.  The
early rapid growth of $E_{tot}$ for times less than ten is due to the 
constant inflow of injected energy with no loss at the high wavenumber
end having yet occurred.  The slower growth for $t > 10$ reflects the
evolution of the spectrum after energy loss begins but is not yet in
equilibrium.  In addition, after $t=10$ the conversion of kinetic energy
into magnetic energy begins to become noticeable at the few percent level.
Throughout all these calculations, the total turbulent energy is always
dominated by the kinetic energy; the maximum magnetic energy integrated
over all scales never exceeds about 70 percent of the kinetic energy.
 
\begin{figure}
\plotone{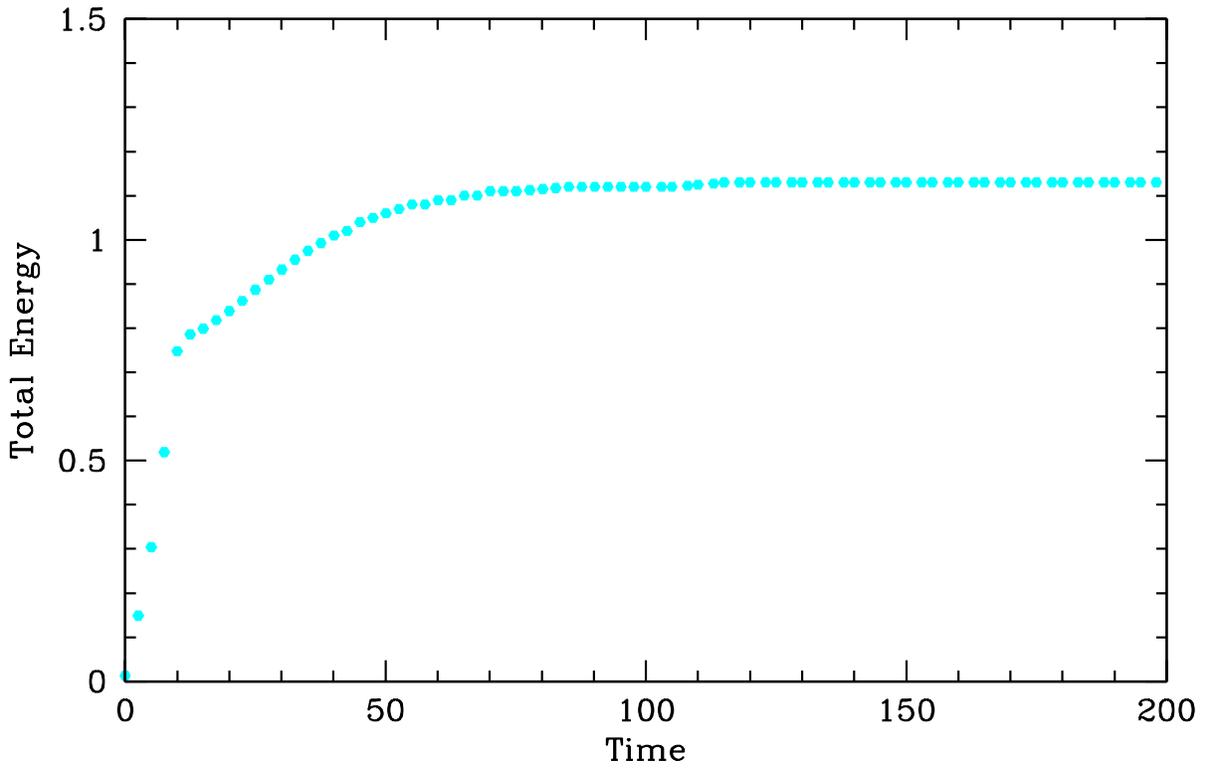}
\caption{Time evolution of the total (kinetic plus magnetic) turbulent energy
for the extended ($k_{top}=1024$) case and delta function energy
injection.  Irregularly spaced data are due to plot roundoff.  
The evolution of $E_{tot}$ is discussed in the text.}
\end{figure}

Additional calculations were made with variations of other parameters 
mentioned in the previous section, such as variation of $F_v$, $\Delta k$,
$\Delta t$, $k_{max}$, and $\nu$.
However, these variations did not significantly alter the overall results
presented in Figures 1 though 4.  In particular, the variation of $F_v$,
which might be thought to have the largest effect, was not particularly
notable.  Calculations were made with values of $F_v$ varying from 0.1 
to 1.0, and while the amplitude of the turbulent energy spectrum increased
as $F_v$ was increased, the shapes of the equilibrium spectra were 
virtually invariant over an order of magnitude change in $F_v$.  This is
not surprising given the tendency for the turbulent MHD cascade to
evolve toward a common power-law form.  Moreover, the scaled time 
required to establish equilibrium was about $t \approx 100$ in all
cases, which is consistent with a scale free calculation.  Variations
in $\Delta t$ and $\nu$ have been discussed above; these are controlled
by the Courant condition to preserve positive definite energies and to
avoid unphysical inverse cascades of energy at large wavenumber.  The
effects of changing the upper limit on the wavenumber range are shown
in the figures; a more clear definition of the loss-free power-law
energy cascade is seen with larger $k_{max}$, but the time required 
to reach equilibrium is not affected.  This is also an expected result,
since equilibrium is reached most rapidly at the smallest scales (largest
$k$).  Decreasing the value of $\Delta k$ and increasing the density of
computational points has the expected effect of making the energy 
transfer among triads of wavenumbers slightly more efficient.  A
doubling of the density of points in $k$ results in maximum change in
$E_v$ of 7 percent at the upper end of the wavenumber range.  Again,
the time to reach equilibrium, which is governed by processes on the
larger scales, is unaffected.

Thus a robust conclusion that emerges from these calculations and which
is illustrated in Figures 2 and 4 is that the turbulent cascade reaches
equilibrium over the wavenumber range $k \in (.01, 1000)$ in about 100
large scale eddy turnover times; $t_{eq} \approx 100 t_{o}$.  This 
result can then be combined with Equation 3 for the time required
after the onset of energy injection to reach turbulent dissipation of
energy into heat;
$$
t_{d} \approx (N\ell_{1}/v)[1 + 1/M + 1/M^{2} + ...+ 1/M^{n-1}] \\
 \approx (N\ell_{1}/v).
$$
Thus these results imply that a value of $N=100$ is appropriate for these
turbulent jet flows.  With this result in hand it is now possible to
estimate the time required for radio AGN outflows to convert their outflow
energy into heat.  To do this requires a conversion from the scale free
calculation here into the relevant astrophysical quantities.

\subsection{Scaling}

The first scale to establish is the spatial scale
that corresponds to $k_{I}$ where the energy is injected.  This is
basically the size of the large scale 
fully developed turbulent eddies which
are driving the turbulence, powered by the vorticity 
generated in the shearing motions of the mixing layer.
As described earlier, the mixing layer grows in thickness until
it permeates the entire jet, at which point the flow is fully
turbulent, and it is at stage that the interior of the jet can
be characterized as a homogeneous and isotropic (in a comoving
frame) turbulent flow.  Hence the size of the large scale
eddies at the point of complete turbulent infiltration 
or saturation is an
appropriate measure of the energy injection scale.  Given the
internal structure of mixing layers, a good approximation
for this eddy size is just the radius of the outflowing jet at
saturation, since that is where the linearly expanding mixing
layers merge together and the potential flow in the core of
the jet ceases.  Experimental data on gaseous turbulent jets
also show that the large scale structures at this point are  
comparable to the jet radius (Dimotakis et al. 1983).
In addition, it is at this point that the
opening angle of the outflow increases slightly and takes
on the common value of $tan (\theta/2) \approx 0.2$ seen in fully
turbulent round jets.  
(Some experimental data for turbulent jets 
also show the creation of larger scale structures that develop
in the downstream flow, but these are often attributed to 
merging or pairing of large eddies formed earlier 
rather than an indication of an energy injection scale 
(e.g., Brown \& Roshko 1974, Dimotakis et al. 1983).
In addition, these structures are most apparent in fluid
turbulent jets and are less dominant in very high Reynolds
number gaseous jets (Dimotakis et al. 1983, Dahm \&
Dimotakis 1990).)  Determination of the jet radius when
it becomes fully turbulent can be estimated from analytic
approximations to $L_{sat}$ or from observations of the
early stages of FR-I outflows.  The latter is probably
the more robust method, given the uncertainty in the 
initial jet densities and outflow speeds. Radio observations
of the inner regions of 3C 31 and 3C 296 (Hardcastle et al.
2002, Laing et al. 2008, Hardcastle et al. 2005) permit an
estimation of the jet width in its earliest turbulent phase.
In particular, the southern jets in these objects show 
an increase of the jet opening angle to
$\sim 25$ degrees that marks the transition to self-similar
fully turbulent outflow.  Estimates of the jet radii at 
these points give values in the the range 
$\approx 0.14-0.17$ kpc, and thus a reasonable estimate
from these data for the scale corresponding to $k_I$ is
about $0.1 - 0.2$ kpc. 

A consistency check can be made
by looking at the corresponding values of $L_{sat}$ and the  
density ratios and outflow speeds that would be required.  
For example, in 3C 31 the estimated point where the jets appear
fully turbulent is at a distance of about 1.7 kpc from the
nucleus, while in the case of 3C 296 this distance is about
$1.5$ kpc.  Using this as a rough measure of $L_{sat}$, and
assuming that the 
functional form for plane mixing layers given above can
be generalized to mixing layers for round jets, given the 
overall similarity of the mixing layers in the two cases
(e.g., Freund et al. 2000,  Dimotakis et al. 1983), 
then it is possible to see if "reasonable"
values of outflow densities and speeds near the nucleus
result.  The jet becomes fully turbulent at a radius $R_{t}
\approx 2 R_{o}$, and using $\eta \approx 0.5$ and $R_{t}  
\approx 0.15$ kpc as an
average of the above numbers, together with the observed
temperatures and inferred number densities for the ambient
medium in 3C 31 and 3C 296 at about 1.5 kpc from the their
nuclei (Hardcastle et al. 2002, 2005) gives jet flow speeds
in the innermost regions of $\sim 10^{9}$ cm s$^{-1}$ and
jet number densities of $\sim 10^{-5}$ cm$^{-3}$,  This is
consistent with models of "light" jets which are mildly
relativistic in the innermost regions 
(e.g., Laing et al. 2008).  
Thus different and independent approaches indicate that a
value of $R_{t} \sim 0.15$ kpc is a consistent value to 
take for the scale of the energy injection range, and this
value will be taken to be the size corresponding to $k_{I}$.

Once a value of $R_{t}$ is established, the remaining scales
can be fixed by setting a velocity scale, and this is usually
chosen to be the turbulent turnover speed of the large scale
eddies in the energy range.  The flow in the mixing layer is
subsonic relative to the ambient medium  
as argued previously, perhaps transsonic at most in
the initial stages.  Hence the value of the eddy turnover
speed $v_{I}$ will be some fraction of the local sound speed
in the ambient medium, and this in turn has been established
for some FR-I objects via observations of x-ray emission in
the central regions of the parent galaxies.  These observations
are consistent with thermal bremsstrahlung radiation, and thus
a temperature and sound speed for this gas can be established.
In particular, Chandra observations of 3C 31 and 3C 296
(Hardcastle et al. 2002, 2005) are consistent with gas
temperatures of order $10^{7}$K, with corresponding sound 
speeds of $\sim 3 \times 10^{7}$ cm s$^{-1}$.  Thus a an
average value of $v_{I} \approx 10^{7}$ in the energy injection
range is consistent with observations and will be
used here.  This value is significantly less than the above estimate
of $10^9$ in the innermost region of the jet, 
and so it is also consistent
with significant deceleration having occurred, especially in the
outer layers of the jet, due to mass entrainment via the growing
turbulent mixing layer.  
Once these scales are established, a time unit
of $t_{o} \approx R_{t}/v_{I}$, the large eddy turnover time,
follows naturally, as does the unit of turbulent kinetic energy density
per unit mass, $v_{I}^{2}$.  This last quantity is the
unit (per unit wavenumber) used in the energy spectra shown 
in Figs. 1-3.

\subsection{Radio AGN Feedback}

The previous two sections provide the information needed to establish 
an estimate of the time required from the establishment of a fully
turbulent outflow in an FR-I jet to the onset of conversion of 
turbulent energy into heat.  As argued above, both
observations and analytic approximations suggest a large scale
eddy size at the energy injection region of $R_{t} \sim 0.15$ kpc,
with an accompanying turbulent eddy speed of $v_{I} \sim 10^{7}$
cm s$^{-1}$.  This gives a time unit for the turbulence
calculations of $t_{o} \approx 4.5 \times 10^{13}$s or $\sim 10^{6}$
yr.  The nonlinear turbulence calculation 
shows the establishment of an equilibrium spectrum
in about 100 large scale eddy turnover times, which gives an estimate of
the time for the onset of conversion of outflow energy into heat
of about 100 million years.  These times are comparable to or less than
the Bremsstrahlung cooling times deduced from x-ray observations of
hot gas in and around the host galaxies of FR-I radio sources
(e.g., Hardcastle et al. 2002, 2005; Laing et al. 2008).

Where is this heat deposited?  The above time estimate is a measure
of the beginning of the conversion into heat; it presumably continues
for a time comparable to the lifetime of the AGN outflow.  This injection of
heat cannot take place within the confines of the outflows that are observed
to be well defined jets, since the energy density in these regions is so
high that conversion of all the directed kinetic energy into heat at that
point would result in pressures high enough to decollimate the jet.  Hence
there is a need to know how far the jets will propagate in the $\sim 10^{8}$
years after they become fully turbulent.  The mean outflow speeds in these
objects are unknown, but an estimate can be gained from the arguments that
the jets are fully turbulent flows.  As mentioned before, such jets are
self-similar flows, and experimental data show that the on-axis
outflow speed in such jets is approximately 10 times the mean outflow speed
on the surface of the jet (e.g., Pope 2000, Hussein et al. 1994).
From previous considerations the outflow speed on the surface of the jet is
of the order of or less than the local sound speed, and this has been
estimated to be $\sim 10^{7}$ cm s$^{-1}$ in the region of turbulent 
energy injection.  Thus in this phase of the jet propagation a mean outflow
speed of $\sim 10^{8}$ cm s$^{-1}$ may be used, which would place the onset
of conversion into heat at a distance of 100 kpc or so from the onset of
the fully turbulent phase.  This is well beyond the collimated outflow region
of most FR-I objects, and hence deposition of heat in this region would not
be in conflict with the observed properties of the jets.  Moreover, this
region may lie in that portion of the outflow where the AGN flow has become
much more broadly dispersed and less well collimated.  This decollimation
process is necessary if radio AGN feedback is to work, since the heating
must take place over a large volume in order to provide significant suppression
of star formation.  
The 100 kpc estimate is somewhat uncertain, since beyond ten
kpc or so from the nucleus the radio morphology of FR-I outflows shows a
clear departure from simple self-similar flow.  
However, it could be that there is
very little deceleration beyond this point, and that
most of the mass entrainment is completed by this time.  The jet morphologies
are not inconsistent with this, in which case the outflow 
speeds near the centerline of the jet remain about the same as they were in
the self-similar regime, and the above estimate of $\sim 100$ kpc
retains some validity.  This is suggestive, but hardly definitive.  There is
in addition the fact that centerline outflow speeds in self-similar jets 
decrease linearly with distance, though the relevant constant in this relation
is unknown; hence there will be some, possibly non-negligible, deceleration of
the jet from the onset of self-similarity until the $\sim 10$ kpc scale is 
reached.  

The late stages of the evolution of these flows raises the issue of their
relation to the AGN outflows seen in the cores of rich galaxy clusters.  
Much work has been done on the role of "radio bubbles" in the ICM, relic and
otherwise, and their possible influence on the evolution of the ICM and 
brightest cluster galaxies (e.g., Fabian et al. 2002, Br\"{u}ggen \& Kaiser 2001,
Fabian et al. 2002, De Young 2003, Jones \& De Young 2005, 
O'Neill et al. 2009), 
and a natural question arises about the relation of these objects to
the flows discussed here.
In general, most FR-I AGN do not show radio "bubbles", nor are they
usually found in the cores of rich clusters.  For those 
that are, it it likely that relatively low powered FR-I jets would be rapidly
decelerated and confined by the high ambient pressures and could slowly inflate
buoyant cavities in the hot ICM, as has been suggested by many models.  Such
cavities are usually small for low radio powers, a few tens of kpc, and the jet
structures feeding them are even smaller (e.g., McNamara \& Nulsen 2007).  
By contrast, the fully turbulent FR-I
jets discussed here are themselves tens of kpc in length before the onset
of complex large scale flow patterns such as bends or meanders. 
Hence there is likely to be a continuum in radio source morphology for a given
radio power as one moves from very high pressure regions in the centers of rich
clusters to lower pressure environments found in small
clusters, groups, and individual galaxies.  The onset of mixing layers, 
entrainment and deceleration could occur in all these environments, as has been
argued here, with the process occurring on smaller scales in more dense and 
high pressure environments (cf. Eq 1.)  
However, one important added factor in the dense ICM in cluster cores is the
presence of non-negligible external 
magnetic fields.  Although the KH instability and
resultant mixing occurs unabated in MHD flows with weak fields (Ryu, Jones \&
Frank 2000), if the average magnetic energy density becomes comparable to the 
kinetic energy density, then the mixing layer picture may have to be 
modified, especially if the external 
fields have long coherence lengths.  In particular,
expansion of an inflated "bubble" into a magnetized medium may not result
in the late stage mixing that simple hydrodynamic flows would suggest
(De Young 2003, De Young et al. 2008), so in the case of cluster cores the
final step in AGN feedback may be more difficult to accomplish.  This issue
is at present still unresolved.

Though the creation of bubbles and cavities may be 
the end state of the flow in high pressure environments, in lower pressure
regions the final state could be much more diffuse and distorted due to 
large scale external magnetic fields or to any 
flows in the circumgalactic gas caused by galaxy orbits or merging events.  The
detection (or not) of such faint extended regions 
is very much a function of x-ray
and radio telescope sensitivity and selection effects.
In this context is it very interesting to note that
Rudnick \& Brown (2009) have recently observed very large, diffuse polarized
radio structures surrounding the FR-I object 3C 31.  Though further observations
are needed, such very large structures could be the signature of the decollimated
and widely dispersed portions of radio AGN outflows where the heat is deposited
from fully turbulent jets.  A similar widely dispersed radio emitting region
has been known for some time to surround the collimated outflows from the FR-I
source associated with M87 (Owen et al. 2000).
Thus the issue of where the turbulent energy is actually deposited remains an
important unresolved problem.  It was argued above that a distance from the
nucleus of $\sim 100$ kpc was not unreasonable for the onset of heat deposition,
but also discussed there were the uncertainties associated with AGN outflow speeds
for distances greater than $\sim 10$ kpc.  For isolated galaxies, heat deposition
at $\sim 100$ kpc is clearly too distant for thermal suppression of star formation
in the near galactic halo, yet the presence of coherent flows on scales $\geq$ 10 kpc
argues for conversion to heat beyond 10 kpc.  The picture is further complicated
by low surface brightness radio data for some FR-I objects (e.g., 3C 296, M87); 
these show the
presence of faint radio emission very close to the stellar extent of the galaxy,
and the implication here is that AGN outflow material from earlier epochs may have
been advected or diffused back into the near circumgalactic regions.  These data
could imply that deposition of heat is possible 
into inner galactic halo at distances a few tens
of kpc from the nucleus, probably on timescales in excess of $10^{8}$ years.
However, it seems clear that if energy deposition is needed for isolated galaxies
in the inner regions of the ISM, further investigation into the very late time
evolution of outflows is needed to see if this process is viable.

\section{Summary and Discussion}

The object of this paper is an exploration of the feasibility of radio
AGN feedback.  The constraints on this process are that it must heat the
interstellar and circumgalactic medium around the host galaxy in a time
less than or the order of the ambient gas cooling time and with enough
volume coverage so that significant star formation can be suppressed and the
host galaxy will appear to be undergoing passive evolution.  The issue of
radio AGN feedback in massive, rich clusters of galaxies is not specifically
addressed, as this process has been treated in many other papers.  
Instead, the more general case of radio AGN feedback for galaxies in
small groups and clusters and even in isolated systems is considered 
here.  The emphasis is to proceed from the basic properties of AGN
outflows to see if a continuous chain of physical processes can be
established that will lead to the required heating in the required time.
The analysis has a small number of fairly simple steps.  First, observations
show that the most common class by far of radio AGN in universe out
to $z \sim 0.5$ and perhaps beyond is the class of FR-I radio sources.
Second, the morphology of these objects strongly suggests
that their bipolar outflows are subsonic or transsonic beyond a few 
kpc from the nucleus.  Third, given the evidence for relativistic
outflows very close to the nucleus together with the nearly universal
onset of global surface instabilities in such shearing flows, this
slow outward motion strongly suggests that it is due to deceleration
of the flow via the formation of non-linear mixing layers and their
accompanying mass entrainment on the surface of the jet.  Fourth,
experimental evidence argues convincingly that such turbulent mixing
layers will soon penetrate throughout the jet volume, resulting in
a fully turbulent flow in the interior of the bipolar outflows.
This conclusion is reinforced by the opening angle of the radio jets
observed in some nearby FR-I objects; this angle is completely
consistent with the experimental data on fully turbulent self-similar
jet outflows.  Fifth, it is well known that such turbulent flows
eventually cascade to smaller and smaller structures until a dissipation
range is reached where most of the turbulent kinetic energy is
converted into heat.  A series of 3D time dependent calculations
of the evolution of such turbulent flows is then performed to see if
this eventual conversion of collimated outflow energy into heat can
take place in a time short enough to be of interest for AGN feedback.
The answer appears to be yes, with the first onset of heating 
occurring roughly $10^{8}$ years after the development of fully
turbulent flow; this time is comparable to or less than the 
cooling times derived from x-ray observations of gas surrounding 
nearby FR-I radio galaxies.  Finally, the question of where this
heat is deposited and whether or not the deposition is sufficiently
isotropic is briefly addressed. To retain the observed radio source
structure, heating probably occurs beyond the extent of the coherent
jet outflow. Estimates of outflow speeds for fully turbulent jets
indicate energy deposition on  
scales as large as $\sim$ 100 kpc, which is well beyond the coherent
jet structures for most FR-I objects.  Isolated galaxies may
require heating on smaller scales, and there is some evidence this 
might be occurring due to late time
evolution of the flows.  
Recent observations of very
large scale and diffuse radio structures around 3C 31 and earlier
observations of M87 suggest that the 
suggested large scale heat deposition
may be taking place. 

These simple arguments, while suggestive, are not conclusive.  It
seems likely that the idea of  
slow and turbulent outflows from FR-I radio AGN
is either almost certainly right or very wrong, 
the latter because we have
missed the fundamental nature of jet outflows from AGN.  For example,
if these flows are "Poynting flows" that are  
completely dominated at all times by magnetic
fields and have very few particles (e.g., Li et al. 2006), then
the nature of the interaction with the surrounding medium might
be very different, with little or no role for MHD or hydrodynamic
turbulence.  However, it is not clear how such magnetically 
dominated outflows can reproduce the morphology observed in the
FR-I radio sources.  Even if turbulent jets are the correct
description, there are details of their evolution suggested by
observations that may mean departures from the isotropic 
turbulence in their interiors that is used here.  Deflections,
bending, or propagation into decreasing density gradients or into
inhomogeneities may all cause departures from this condition and
perhaps changes in the time required for turbulence to decay
into heat.  The magnitude of such departures is currently unknown,
and they may be small enough that the above treatment remains
a good approximation, especially in the interior of the jet.
Finally, the large scale distribution of
energy from radio AGN outflows remains a largely unsolved
problem.  The observational hints are suggestive, 
but more observations of faint and
diffuse radio emission around FR-I outflows are needed, as are
additional calculations of the very late stages of these
outflows.

\acknowledgments

Portions of this work were completed during the Aspen Center for Physics astrophysics
programs in 2008 and 2009.  Thanks are due to Bob Rosner and to Tom Jones for illuminating
discussions and to Jessica Moy for bibliographic assistance.

\appendix
\section{Appendix: The EDQNM Approximation}

Presented here are some additional concepts and relations that lie
behind the derivation of Equation (2).  Most of the basic framework
of the EDQNM method is given in the main body of the paper, and 
complete derivations for all aspects of the method, together with
more detailed descriptions of the method and its variants, are
provided in the references cited.  A particularly complete treatment
is found in Lesieur (2008).

The foundation of the EDQNM model are the familiar conservation equations
of ideal MHD flow:

\begin{equation}
(\partial/\partial t - \nu\nabla^{2}){\bf v} = -({\bf v \cdot\nabla)v} + 
({\bf b\cdot\nabla)b} - \nabla p + {\bf f}
\end{equation}
\begin{equation}
(\partial/\partial t - \lambda\nabla^{2}){\bf b} = -({\bf v\cdot\nabla)b + (b\cdot\nabla)v}
\end{equation}
\begin{equation}
{\bf \nabla\cdot b} = 0; {\bf \nabla\cdot v} = 0
\end{equation}
 where ${\bf B(r},t) = (\rho\mu)^{1/2}{\bf b(r},t)$ is the magnetic field,
$\rho$ is the density, ${\bf v(r},t)$ is the fluid velocity, $\rho p$ is
the pressure and $\mu$ is the magnetic susceptibility.  Here $\nu$ is 
the kinematic viscosity, $\lambda$ is the magnetic diffusivity, and 
${\bf f}$ is a forcing term that represents the injection of kinetic
energy into the system.  In principal the formalism also accommodates
the injection of magnetic energy as well as kinetic and magnetic
helicities (${\bf v\cdot\nabla \times v}$ and ${\bf b\cdot\nabla \times
b}$).  The treatment described here is restricted in nearly incompressible flows
that are almost isotropic in that they admit helicity in the flow fields.  
Extensions of the method to anisotropic and quasi compressible flows are 
described in the references.

As mentioned in the text, the EDQNM method derives equations that govern the 
second order moments of velocity and magnetic field which thus provide the evolution
of the kinetic and magnetic energy densities.  These governing equations actually
determine the behavior of the Fourier transformed moments; because Fourier components
are global averages over configuration space, the equations provide the evolution of
energy on a particular scale over the entire flow and thus provide immediately the
time evolution of the energy spectra.  In addition, operation in Fourier space allows
in this formalism the integration of the governing equations using logarithmic intervals
in wavenumber space and so permits coverage over many orders of magnitude in spatial
scale in three dimensions, a task that would be extremely difficult through direct 
numerical simulation.  The general form of Equation (2) in the text can be obtained
by first taking the Fourier transform of the above Navier-Stokes equation, which gives,
for no magnetic fields and no helicity, with $\rho$ explicit, the general form
$$
(\partial/\partial t + \nu k^{2})u (k,t) = F + F.T.[(1/\rho){\bf \nabla}p +
{\bf u \cdot \nabla u}]
$$
where $u$ is the Fourier transform of the velocity $v$, 
$F$ is the Fourier transform of the forcing (energy injection) term and F.T.
is the Fourier transform operator.  The transformed incompressibility condition
becomes ${\bf k \cdot u (k},t) = 0$, and this condition allows elimination of the
pressure from the equation of motion.  This is because 
the incompressibility equation shows that
${\bf u}$ is in a plane perpendicular to ${\bf k}$, and thus so are $\partial {\bf 
u}/\partial t$ 
and $\nu k^{2} {\bf u}$, while the pressure gradient $i p {\bf k}$ is
parallel to ${\bf k}$.  
As a result, the F.T. of ${\bf v \cdot \nabla v}
+ (1/\rho){\bf \nabla}p$ is the projection onto the plane containing ${\bf u}$ of
the F.T. of the non-linear term ${\bf v \cdot \nabla v}$.  Thus the form of the fluid
equation in Fourier space becomes, in general,
\begin{equation}
(\partial/\partial t + \nu k^{2})u (k,t) = F + ikG(k) \int u (p,t)
u (q,t)dp.
\end{equation}
where the vector notation has been dropped.  The exact form of this equation can be
found in Lesieur (2008).  It can now be seen that a similar equation can be found for
the Fourier transform of the second moment of Eq. A1, and it is also clear that the
result of this will include a third order term in $u$ on the right hand side. Similarly
an equation for the third order moments will include an integral over a term of fourth
order moments; this difficulty is of course a result of the non-linear nature of the
flow arising from the $v \cdot \nabla v$ term in the momentum equation.  Closure of this
system of moments is accomplished by assuming that the fourth order moments are nearly
Gaussian; hence the "quasi-normal" approximation.  This approximation allows the fourth
order moments to be written as sums of products of the second order moments, and 
because the third order moments can already be written as a function of second order
moments, the system is closed at this level.  
A discussion of the accuracy of this approximation 
and tests of its rigor are found in Lesieur (2008) and references therein.  Thus the
origin of the overall form of Eq (2) in the text can be seen. In general the integral
in this equation will contain many terms that are quadratic in kinetic energy, helicity,
and magnetic energy, and numerical solution of this integro-differential equation is
necessary.  It can be shown (Lesieur 2008) that energy is strictly conserved when 
integration over wavenumber space includes all triads of turbulent eddies whose 
wavenumbers (k,p,q) satisfy the requirement that ${\bf k + p + q = 0}$; thus interactions
with $k \ll p \sim q$ include non-local interactions in wavenumber space. 
A complete tabulation of all the terms included on the right side of Eq (1)
is found in De Young (1980).  

In addition to a closure approximation, two other constraints are added to the right
side of Eq. (2).  In its simplest form, the quasi-normal approximation can,
at late times, lead to negative energies in some portions of the spectra.  This
has been shown to be due to an eventual excess value building up in the 
third order moments.  These moments have been shown experimentally to
saturate, and a modification to the theory incorporates this effect by adding
a linear damping term in the equation governing the growth of the third order
moments which represents the deformation rate of
eddies of size $\sim k^{-1}$ by larger scale eddies.  This "eddy damping"
stabilizes the growth rate of the third order moments and results in the 
Eddy Damped Quasi-Normal (EDQN) version of the theory.  A final modification
guarantees the positive definite character of the spectra in all situations,
not just at late times.  This is the introduction of a Markov process to 
remove excessive "memory" in the system on timescales of order of the large
eddy turnover time, including removal of any time variations in the eddy
damping term.  With this step the formalism takes on the final form
of the EDQNM theory.  Specific forms for these modifications can be found
in De Young (1980) and Lesieur (2008).


\clearpage



\end{document}